\documentclass[prd,%preprint,
,superscriptaddress,%showpacs,
nofootinbib,%
tightenlines ]{revtex4}
\usepackage{epsfig}
\usepackage[colorlinks=true,linkcolor=blue,urlcolor=blue,citecolor=blue]{hyperref}
\usepackage{amsfonts}
\usepackage{amssymb}
\usepackage{amsmath}
\usepackage{slashed}

\newcommand{\ben}{\begin{displaymath}}
\newcommand{\een}{\end{displaymath}}
\newcommand{\be}{\begin{equation}}
\newcommand{\ee}{\end{equation}}
\newcommand{\bea}{\begin{eqnarray}}
\newcommand{\eea}{\end{eqnarray}}

\begin{document}
\title{Gravitational form factors of the delta resonance in chiral EFT}
\author{H.~Alharazin}
\affiliation{Institut f\"ur Theoretische Physik II, Ruhr-Universit\"at Bochum,  D-44780 Bochum,
 Germany}
\author{E.~Epelbaum}
 \affiliation{Institut f\"ur Theoretische Physik II, Ruhr-Universit\"at Bochum,  D-44780 Bochum,
 Germany}
\author{J.~Gegelia}
 \affiliation{Institut f\"ur Theoretische Physik II, Ruhr-Universit\"at Bochum,  D-44780 Bochum,
 Germany}
 \affiliation{High Energy Physics Institute, Tbilisi State
University, 0186 Tbilisi, Georgia}
\author{U.-G.~Mei\ss ner}
 \affiliation{Helmholtz Institut f\"ur Strahlen- und Kernphysik and Bethe
   Center for Theoretical Physics, Universit\"at Bonn, D-53115 Bonn, Germany}
 \affiliation{Institute for Advanced Simulation, Institut f\"ur Kernphysik
   and J\"ulich Center for Hadron Physics, Forschungszentrum J\"ulich, D-52425 J\"ulich,
Germany}
\affiliation{Tbilisi State  University,  0186 Tbilisi,
 Georgia}
\author{B.-D. Sun}
\affiliation{Guangdong Provincial Key Laboratory of Nuclear Science,
Institute of Quantum Matter, \\South China Normal University, Guangzhou 510006, China} 
\affiliation{Guangdong-Hong Kong Joint Laboratory of Quantum Matter,\\
Southern Nuclear Science Computing Center, \\South China Normal University, Guangzhou 510006, China
}
\affiliation{Helmholtz Institut f\"ur Strahlen- und Kernphysik and Bethe
   Center for Theoretical Physics, Universit\"at Bonn, D-53115 Bonn, Germany}

\begin{abstract}
The leading one-loop corrections to the gravitational form factors of the delta resonance
are calculated in the framework of chiral effective field theory. 
Various contributions to the energy-momentum tensor and the
renormalization of the low-energy constants are worked out.  
Using the small scale expansion, expressions for static quantities are obtained and
the real and imaginary parts of the gravitational form factors are calculated numerically.
\end{abstract}

\maketitle

\section{Introduction}

The linear response of the effective action to the change of the space-time metric
specifies mechanical properties of particles with various spins.
Static characteristics, like the mass, spin and the $D$-term correspond to the hadron
gravitational form factors (GFFs) at zero momentum transfer \cite{Kobzarev:1962wt,Pagels:1966zza}.
Determining the third mechanical characteristics (the $D$-term)
of a particle is a more difficult problem than the  mass and spin, which are well-studied
and well-measured quantities. The $D$-term of the nucleon has been related 
to the distribution of the internal  forces in Ref.~\cite{Polyakov:2002yz} (for a  review
see e.g. Ref.~\cite{Polyakov:2018zvc}).  
%However its value is not solely determined by the strong interaction, but may also be affected by gravitational interaction. 
Poincar\'e covariance of the hadron states guarantees that the leading two GFFs at zero
momentum-transfer correspond to mass and spin~\cite{Lorce:2019sbq}. 
Adding total derivatives to the energy-momentum tensor (EMT) leaves the Poincar\'e group
generators unaffected, i.e.~it does not impact the particle's mass or spin. Therefore,
these static characteristics are well constrained.
However, Poincar\'e symmetry does not protect the $D$-term. By adding total derivatives
to the EMT, the $D$-term can be significantly changed.
For example, %for spin-0 particles,
it has been shown in Ref.~\cite{Hudson:2017xug} that even infinitesimally small interactions
(minimally coupled to gravity) can drastically impact the $D$-term: while the free Klein-Gordon
theory predicts the $D$-term of a spin-0 particle to be $D=-1$, it becomes $D=-1/3$ after
inserting an ``improvement term'' in the EMT. The improvement term is obtained by requiring
that the action respects the conformal symmetry of the classical theory in the massless limit,
which is equivalent to adding a term of non-minimal coupling $-\frac{1}{12}R\Phi^2$ to the
action (where $R$ is the Riemann scalar, and $\Phi$ is the free Klein-Gordon field). On a
quantum level, the conformal symmetry is broken, but the improvement term is required to
remove UV divergences up to three loops in dimensional regularization in the
Klein-Gordon theory~\cite{Collins:1976vm}.

In recent years, GFFs have attracted increasing attention for characterizing properties of
hadrons with different spins due to their connection to generalized parton distributions (GPDs). 
Parameterizations of the EMT matrix elements in terms of GFFs for hadrons have been considered
for spin-0~\cite{Pagels:1966zza}, spin-1~\cite{Holstein:2006ud,Cosyn:2019aio,Polyakov:2019lbq},
and for arbitrary-spin particles~\cite{Cotogno:2019vjb}. 
The mechanical properties, energy and spin densities as well as spatial distributions of
pressure and shear forces have been introduced for spin-0 and spin-1/2 in
Ref.~\cite{Polyakov:2002yz}, 
and later also for systems with higher spins
\cite{Polyakov:2019lbq,Panteleeva:2020ejw,Kim:2020lrs}.

The nucleon gravitational form factors can be measured experimentally 
in exclusive processes like deeply virtual Compton scattering  (DVCS)
\cite{Ji:1996ek,Radyushkin:1997ki} and hard exclusive meson production \cite{Collins:1996fb}.
The connection to GFFs can be seen in the QCD description of these processes, where the
{\it symmetric} energy-momentum tensor (EMT)   appears naturally in the operator product expansion,
see derivation, e.g., in Ref.~\cite{Ji:1996ek}.  
The first  results of  measurements of the $D$-term
in hard QCD processes became available  in Refs.~\cite{Kumericki:2015lhb,Nature} for the nucleon,
and in Ref.~\cite{Kumano:2017lhr} for the pion. Profound studies of 
subtleties in the extraction of the $D$-term  from hard exclusive processes can be found
in Ref.~\cite{Kumericki:2019ddg}.
The GFFs  have also been studied in lattice QCD, see e.g.
Refs.~\cite{Shanahan:2018nnv,Shanahan:2018pib,Alexandrou:2013joa,Bratt:2010jn,Hagler:2007xi}
and references therein.

In contrast to the electromagnetic properties of the delta  resonances, which have been
extensively investigated both experimentally and theoretically~\cite{Pascalutsa:2006up,Fu:2022rkn},
it is very difficult to measure experimentally the GFFs of the $\Delta$ or to extract them
from the corresponding GPDs because of the short-lived nature of the $\Delta$ resonance.
Several calculations have been done to estimate these GFFs. For example, the SU(2)
Skyrme model~\cite{Kim:2020lrs}, the quark-diquark model~\cite{Fu:2022rkn} and a
lattice QCD approach (for the gluonic part)~\cite{Pefkou:2021fni} have been employed
to calculate the  GFFs of the spin-3/2 delta resonances 
and the corresponding pressure and shear forces. 
%More references on other approaches for spin-0 and spin-1/2 cases can also found in Refs.~\cite{Kim:2020lrs,Fu:2022rkn} and a review by Ref.~\cite{Polyakov:2018zvc}. 
Except for the constraints that the mass and spin should be obtained from the zero
momentum-transfer limits of the leading two GFFs, the other GFFs calculated in
Refs.~\cite{Kim:2020lrs} and~\cite{Fu:2022rkn} show notable differences and even different
signs for the $D$-term ($D=-3.53$ in Ref.~\cite{Kim:2020lrs} and $D=0.986$ in
Ref.~\cite{Fu:2022rkn}).  Notice that the negative sign of the $D$-term is expected to
follow from the stability condition.  More systematic studies are thus required to
investigate the properties of the GFFs of delta resonance, e.g., to what extent the expected
stability condition holds, taking into account that the deltas are unstable.

For systematic studies of  low-energy hadronic processes with delta resonances
induced by gravity one may rely on the effective chiral Lagrangian for nucleons, pions and
delta resonances in  curved spacetime.
Effective Lagrangian of pions in curved spacetime has been derived in Ref.~\cite{Donoghue:1991qv}, 
and the GFFs of the pion can be found in Ref.~\cite{Kubis:1999db}. 
The effective chiral Lagrangian of order two for nucleons and pions in curved spacetime,
along with the calculation of the leading one-loop 
contributions to the nucleon GFFs, can be found in Ref.~\cite{Alharazin:2020yjv}. 
In this work we apply chiral effective field theory (EFT) to calculate the one-particle
matrix elements of EMT for delta resonances within the EOMS scheme to be discussed below.

Our paper is organized as follows: In section~\ref{effective_Lagrangian} we specify the
relevant terms of the  effective Lagrangian for pions, nucleons and delta resonances in
curved spacetime  and the corresponding expression for the EMT.
We calculate the gravitational form factors of the delta resonance in section~\ref{NFFs}.  
%The large distance asymptotic of the energy, spin, pressure and shear force distributions is studied in subsection~\ref{sec:largeR}.
The results of our work are summarized in  section~\ref{conclusions}.

\section{Effective action in curved space time and the energy-momentum tensor}
\label{effective_Lagrangian}

The action corresponding to the leading-order contributions to the effective Lagrangian of
pions, nucleons and delta resonances interacting with an external gravitational field can
be straightforwardly obtained from the corresponding expressions in the flat spacetime.
It has the following form:
\begin{eqnarray}
S_{\rm \pi}^{(2)} &=& \int d^4x \sqrt{-g}\, \left\{ \frac {F^2}{4}\, g^{\mu\nu}\, {\rm Tr}
( D_\mu U  (D_\nu U)^\dagger ) + \frac{F^2}{4}\,{\rm Tr}(\chi U^\dagger +U \chi^\dagger) \right\}\,,
\label{PionAction}
\\
S_{\rm \pi N}^{(1)}  & = & \int d^4x \sqrt{-g}\, \biggl\{\, \bar\Psi \, i  \gamma^\mu
\overset{\leftrightarrow}{\nabla}_\mu \Psi -m \bar\Psi\Psi +\frac{g_A}{2}\, \bar\Psi \gamma^\mu
\gamma_5 u_\mu \Psi  \biggr\} \,, \label{PiNAction} \\
S_{\pi \Delta}^{(1)} & = & - \int d^4 x  \sqrt{-g} \biggl[ g^{\mu\nu} \, 
\Bar{\Psi}^{i}_\mu  \,  i \gamma^\alpha \overset{\leftrightarrow}{\nabla}_\alpha  \Psi^{i}_\nu  -
m_\Delta \, g^{\mu\nu}  \Bar{\Psi}^{i}_\mu   \Psi^{i}_\nu  -  g^{\lambda\sigma} \left( \Bar{\Psi}^{i}_\mu
i \gamma^{\mu}{\overset{\leftrightarrow}{\nabla}_\lambda}  \Psi^{i}_\sigma   +  \Bar{\Psi}^{i}_\lambda
i \gamma^{\mu}{\overset{\leftrightarrow}{\nabla}_\sigma}  \Psi^{i}_\mu  \right)  
\nonumber\\
&+&   i  \Bar{\Psi}^{i}_\mu \gamma^\mu \gamma^\alpha\gamma^\nu \overset{\leftrightarrow}{\nabla}_\alpha
\Psi^{i}_\nu + m_\Delta \Bar{\Psi}^{i}_\mu \gamma^\mu \gamma^\nu  \Psi^{i}_\nu  +   \frac{g_1}{2}
\,g^{\mu\nu}\bar{\Psi}^i_{\mu} u_\alpha \gamma^\alpha \gamma_5 \Psi^i_{\nu} 
+  \frac{g_2}{2} \bar{\Psi}^i_{\mu}  \left( u^\mu  \gamma^\nu + u^\nu \gamma^\mu \right) \gamma_5
\Psi^i_{\nu}     \nonumber \\
&  + & \frac{g_3}{2} \bar{\Psi}^i_{\mu}   u_\alpha  \gamma^\mu \gamma^\alpha \gamma_5  \gamma^\nu
\Psi^i_{\nu} \biggr]\,,     
\label{MAction}
\end{eqnarray}
where the delta resonance is represented by the Rarita-Schwinger field. Note that the delta
fields $\Psi^\mu_i$ contain isospin projectors $\xi_{ij}^{\frac{3}{2}}=\delta_{ij}-\frac{1}{3}\tau_{i}
\tau_{j}$, i.e. they satisfy the condition  $\Psi^\mu_i = \xi_{ij}^{\frac{3}{2}} \Psi^\mu_j $.
Here,  $\mu$ and $i$ are the Lorenz and isospin-indices, $g^{\mu\nu}$ is the metric with the
signature $(+,-,-,-)$ and $\gamma_\mu \equiv e_\mu^a \gamma_a $, with $ e_\mu^a$ denoting
the vielbein gravitational fields. It follows from the consistency conditions, imposed
on the Lagrangian with Rarita-Schwinger fields, that the following relations for the
low-energy constants $g_1$, $g_2$ and $g_3$ must be satisfied: 
$g_2 = g_3 = -g_1$ \cite{Wies:2006rv}.\footnote{In this work we take the off-shell
parameter $A$ equal to $-1$, which removes the dependence on the spacetime dimension in the
action $S_{\pi \Delta}^{(1)}$ in Eq.(\ref{MAction}) and simplifies the free propagator of the
delta resonance. See also the discussion in Ref.~\cite{Krebs:2009bf}.}
Further, $m$, $g_A$ and $F$ refer to nucleon mass, the axial vector coupling of the nucleon and
the pion decay constant in the chiral limit, respectively (also called bare parameters later). 
The action corresponding to the leading-order chiral Lagrangian containing pions, nucleons
and deltas interacting with an external gravitational field is %\cite{Hacker:2005fh}
 \begin{eqnarray}
S_{\pi N\Delta}^{(1)}&=&-\int d^4 x  \sqrt{-g} \; g_{\pi N\Delta}~\bar{\Psi}_{\mu,i}  \left( g^{\mu \nu}
- \gamma^\mu \gamma^\nu \right)  u_{\nu,i}\Psi+\text{H.c.},
\label{piND}
 \end{eqnarray}
where $u_\mu=\tau_i u_{\mu,i}$. The pion-field dependent matrix $u_\mu$ will be specified below.
From the second-order chiral Lagrangian containing pions and deltas 
interacting with an external gravitational field we need to consider the following term
\cite{Yao:2016vbz}:
\begin{eqnarray}
S_{\pi \Delta, a}^{(2)}  & = & \int d^4 x  \sqrt{-g}\,a_1\, \bar{\Psi}^i_{\mu} \Theta^{\mu \alpha}(z)
\left< \chi_+ \right> g_{\alpha \beta}  \Theta^{\beta \nu}(z^\prime)\Psi^i_{\nu} \,,
\label{MAction2} 
 \end{eqnarray}
where $z$ and $z^\prime$ are two independent parameters. Further terms of the second order
Lagrangian contributing (at tree order) to our calculations of the GFFs of delta resonance
contain only the Riemann curvature tensor, the Ricci tensor and the Ricci scalar.
The most general second-order Lagrangian of such terms involves either the curvature scalar or
the Riemann and Ricci tensors and reduces to the following minimal form: 
\begin{eqnarray}
  S_{\pi\Delta,b}^{(2)}  & = & \int d^4 x  \sqrt{-g} \biggl[ h_1 R~g^{\alpha\beta} \bar{\Psi}^i_\alpha\Psi_\beta^i + h_2 R~\bar{\Psi}^i_\alpha \gamma^{\alpha} \gamma^{\beta} \Psi_\beta^i +i  h_3 R\left( g^{\alpha \lambda} \bar{\Psi}^i_\alpha \gamma^{\beta}\overset{\rightarrow}{\nabla}_\lambda \Psi_\beta^i- g^{\beta \lambda} \bar{\Psi}^i_\alpha  \gamma^{\alpha}\overset{\leftarrow}{\nabla}_\lambda \Psi_\beta^i \right)  \nonumber \\ &+& h_4 R^{\mu\nu}~\bar{\Psi}^i_\mu\Psi_\nu^i +   2 i h_5 R^{\mu\nu}~g^{\alpha\beta}  \bar{\Psi}^i_\alpha \gamma_\mu \overset{\leftrightarrow}{\nabla}_\nu\Psi_\beta^i +i h_6 R^{\mu\nu} g^{\alpha \beta}  \left(\bar{\Psi}^i_\alpha  \gamma_\mu \overset{\rightarrow}{\nabla}_\beta\Psi_\nu^i  -\bar{\Psi}^i_\nu  \gamma_\mu \overset{\leftarrow}{\nabla}_\beta\Psi_\alpha^i  \right)
\nonumber \\ &+&  i h_7 R^{\mu\nu} \left(\bar{\Psi}^i_\alpha  \gamma^\alpha \overset{\rightarrow}{\nabla}_\mu\Psi_\nu^i  -\bar{\Psi}^i_\nu  \gamma^\alpha \overset{\leftarrow}{\nabla}_\mu\Psi_\alpha^i  \right) + h_8 R^{\mu\nu} \left( \bar{\Psi}^i_\alpha  \gamma^\alpha\gamma_\mu \Psi_\nu^i  + \bar{\Psi}^i_\nu  \gamma_\mu\gamma^\alpha \Psi^i_\alpha   \right) \nonumber \\ &+& i  h_9 R^{\mu\nu} \left( \bar{\Psi}^i_\kappa \gamma^\kappa \gamma^\alpha\gamma_\mu \overset{\rightarrow}{\nabla}_\nu \Psi_\alpha^i  - \bar{\Psi}^i_\alpha \gamma_\mu \gamma^\alpha  \gamma^\kappa \overset{\leftarrow}{\nabla}_\nu \Psi^i_\kappa   \right) + i h_{10} R^{\mu\nu\alpha\beta} \bar{\Psi}^i_\alpha \sigma_{\mu\nu} \Psi_\beta^i  \nonumber\\ &+&  i \left[ h_{11}~R^{\mu\nu\alpha\beta}+ h_{12}~R^{\mu\alpha\nu\beta} \right] \left(\bar{\Psi}^i_\alpha  \gamma_\mu \overset{\rightarrow}{\nabla}_\nu\Psi_\beta^i  -\bar{\Psi}^i_\beta  \gamma_\mu \overset{\leftarrow}{\nabla}_\nu\Psi_\alpha^i  \right) + h_{13} R^{\mu\alpha\nu\beta} \bar{\Psi}^i_\alpha  \gamma_\mu\gamma_\nu \Psi_\beta^i \nonumber \\ &+&  i \left[ h_{14}~R^{\mu\nu\alpha\beta}+ h_{15}~R^{\mu\alpha\nu\beta} \right] \left( \bar{\Psi}^i_\kappa  \gamma^\kappa\gamma_\mu \gamma_\nu \overset{\rightarrow}{\nabla}_\alpha \Psi_\beta^i   -  \bar{\Psi}^i_\beta  \gamma_\nu \gamma_\mu \gamma^\kappa \overset{\leftarrow}{\nabla}_\alpha \Psi_\kappa^i  \right) \biggr]\,, 
 \label{MActionh}
  \end{eqnarray}
where the $h_i$ are coupling constants.

The various building blocks of the effective Lagrangian are defined as follows:
\begin{eqnarray}
D_\mu U & = & \partial_\mu U -i r_\mu U +i U l_\mu \,,\nonumber\\ \overset{\leftrightarrow}{\nabla}_\mu &= &\frac{1}{2}( \overset{\rightarrow}{\nabla}_\mu - \overset{\leftarrow}{\nabla}_\mu ), \nonumber\\ 
 \Theta^{\mu \nu}(z) &=& g^{\mu\nu} + z \gamma^{\mu} \gamma^\nu ,\nonumber \\
 \overset{\rightarrow}{\nabla}_\mu \Psi^i_\nu & = &  \nabla_\mu^{ij} \Psi^j_\nu = \left[
    \delta^{ij}\partial_\mu + \delta^{ij}\Gamma_\mu-i\delta^{ij} v_\mu^{(s)}-i \epsilon^{ijk}\text{Tr}\left(\tau^k\Gamma_\mu\right) 
    + \frac{i}{2}\delta^{ij}\omega^{ab}_\mu \sigma_{ab}\right]\Psi^j_\nu - \Gamma^{\alpha}_{\mu\nu}\Psi^i_\alpha, \nonumber \\ 
    \bar\Psi^i_\nu \overset{\leftarrow}{\nabla}_\mu  & = & \nabla_\mu^{ij} \Psi^j_\nu =  \bar\Psi^j_\nu \left[
    \delta^{ij}\partial_\mu - \delta^{ij}\Gamma_\mu + i\delta^{ij} v_\mu^{(s)} + i \epsilon^{ijk}\text{Tr}\left(\tau^k\Gamma_\mu\right) 
    - \frac{i}{2}\delta^{ij}\omega^{ab}_\mu \sigma_{ab}\right] - \bar\Psi^i_\alpha\Gamma^{\alpha}_{\mu\nu}, \nonumber \\ 
    \overset{\rightarrow}{\nabla}_\mu \Psi 
    & = & \partial_\mu\Psi +\frac{i}{2} \, \omega^{ab}_\mu \sigma_{ab} \Psi + \left( \Gamma_\mu  -i v_\mu^{(s)}\right)\Psi\,,
    \nonumber\\
\bar\Psi \overset{\leftarrow}{\nabla}_\mu & = & \partial_\mu\bar\Psi -\frac{i}{2} \, \bar\Psi \, \sigma_{ab} \, \omega^{ab}_\mu - \bar\Psi \left( \Gamma_\mu  -i v_\mu^{(s)}\right) \,, \nonumber\\
u_\mu & = & i \left[ u^\dagger \partial_\mu u  - u \partial_\mu u^\dagger -i (u^\dagger v_\mu u - u v_\mu u^\dagger )\right]\,,\nonumber\\
\chi & = & 2 B_0(s+i p)\,,\nonumber\\
\Gamma_\mu & = & \frac{1}{2} \left[ u^\dagger \partial_\mu u  +u \partial_\mu u^\dagger -i (u^\dagger v_\mu u+u v_\mu u^\dagger )\right]~,\nonumber\\
\omega_\mu^{ab} &=& -\frac{1}{2} \, g^{\nu\lambda} e^a_\lambda \left( \partial_\mu e_\nu^b
- e^b_\sigma \Gamma^\sigma_{\mu \nu} \right),\nonumber\\
\Gamma^\lambda_{\alpha \beta} &=& \frac{1}{2}\,g^{\lambda\sigma} \left( \partial_\alpha g_{\beta\sigma}
+ \partial_\beta g_{\alpha\sigma} -  \partial_\sigma g_{\alpha\beta} \right)~, \nonumber\\
R^\rho_{~\sigma\mu\nu} &=& \partial_\mu \Gamma^\rho_{\nu \sigma} -\partial_\nu \Gamma^\rho_{\mu \sigma} + \Gamma^\rho_{\mu \lambda}  \Gamma^\lambda_{\nu \sigma} - \Gamma^\rho_{\nu \lambda}  \Gamma^\lambda_{\mu \sigma}   \,,\nonumber\\ R_{\mu \nu} &=& R^\lambda_{~\mu\lambda\nu} \,,\nonumber\\ 
R &=& g^{\mu\nu} R^\lambda_{~\mu\lambda\nu},\, \nonumber\\
\sigma_{\mu\nu} &= &\frac{i}{2}\left[\gamma_{\mu},\gamma_{\nu}\right]\,, \nonumber\\
 \chi_+ &=& u^\dagger  \chi u^\dagger  + u   \chi^\dagger u, 
\label{CovD}
\end{eqnarray}
where  %$\sigma_{ab}=\frac{i}{2}[\gamma_a, \gamma_b]$, 
the $2\times 2$ unitary matrix $U$ represents the pion field,  $s$, $p$, $l_\mu =v_\mu-a_\mu $,
$r_\mu =v_\mu + a_\mu $ and $v_\mu^{(s)}$ refer to the external sources, $\chi= 2 B_0(s+i p)$, 
and the parameter $B_0$ is related to the  vacuum condensate in the chiral limit.  
The vielbein fields satisfy the following relations:
\begin{eqnarray}
&& e^a_\mu e^b_\nu \eta_{ab}=g_{\mu\nu}, \ \ \  e_a^\mu e_b^\nu \eta^{ab}=g^{\mu\nu}, \nonumber\\
&& e^a_\mu e^b_\nu g^{\mu\nu}=\eta^{ab}, \ \ \  e_a^\mu e_b^\nu g_{\mu\nu}=\eta_{ab}.
\label{VRels}
\end{eqnarray}

%\medskip

Using the definition of the EMT for matter fields interacting with the gravitational metric field,
\begin{eqnarray}
T_{\mu\nu} (g,\psi) & = & \frac{2}{\sqrt{-g}}\frac{\delta S_{\rm m} }{\delta g^{\mu\nu}}\,,
\label{EMTMatter}
\end{eqnarray}
we obtain  { in flat spacetime} from the action of Eq.~(\ref{PionAction}):
\begin{eqnarray}
T^{{(2)}}_{\pi , \mu\nu}  & = &  \frac {F^2}{4}\, {\rm Tr} ( D_\mu U  (D_\nu U)^\dagger)
-\frac{ \eta_{\mu\nu}}{2} \left\{ \frac {F^2}{4}\, {\rm Tr} ( D^\alpha U  (D_\alpha U)^\dagger )
+  \frac{F^2}{4}\,{\rm Tr}(\chi U^\dagger +U \chi^\dagger) \right\} + \left( \mu \leftrightarrow
\nu \right),
\label{PionEMT}
\end{eqnarray}
where $\eta_{\mu\nu}$ is the Minkowski metric tensor. 
For the fermion fields interacting with the gravitational vielbein fields we use the definition
\cite{Birrell:1982ix} 
\begin{eqnarray}
T_{\mu\nu}  (g,\psi) & = & \frac{1}{2 e} \left[ \frac{\delta S }{\delta e^{a \mu}} \,e^{a}_\nu
+ \frac{\delta S }{\delta e^{a \nu}} \,e^{a}_\mu  \right] ,
\label{EMTfermion}
\end{eqnarray}
where $e$ denotes the determinant of $e^a_\mu$. 
The action of Eq.~(\ref{PiNAction}) leads to the following expression for the EMT  in
flat spacetime:
\begin{eqnarray}
T^{{  (1)}}_{\pi N, \mu\nu}  & = &  \frac{i}{2} \bar\Psi \,  \gamma_\mu  \overset{\leftrightarrow}{D}_\nu \Psi + \frac{g_A}{4}  \bar\Psi \,  \gamma_\mu \gamma_5 u_\nu \Psi - \frac{\eta_{\mu\nu}}{2}\biggl(\, \bar\Psi \, i  \gamma^\alpha \overset{\leftrightarrow}{D}_\alpha \Psi -m \bar\Psi\Psi +\frac{g_A}{2}\, \bar\Psi \gamma^\alpha \gamma_5 u_\alpha \Psi \biggr) + \left( \mu \leftrightarrow \nu \right)   \,,
\label{MEMT}
\end{eqnarray}

The actions of Eqs.~(\ref{MAction}), (\ref{piND}) and (\ref{MAction2}) lead to the following expressions for the EMT  in flat spacetime:%\footnote{We use here the notation:%.}: 
\begin{eqnarray}
  T^{{  (1)}}_{\pi \Delta,\mu\nu} & = &   -\Bar{\Psi}^{i}_\mu  \,  i  \gamma^\alpha \overset{\leftrightarrow}{D}_\alpha  \Psi^{i}_\nu + \Bar{\Psi}^{i}_\alpha  \,  i  \gamma^\alpha \overset{\leftrightarrow}{D}_\mu  \Psi^{i}_\nu   +
       \Bar{\Psi}^{i}_\mu  \,  i  \gamma^\alpha \overset{\leftrightarrow}{D}_\nu  \Psi^{i}_\alpha + m_\Delta \Bar{\Psi}^{i}_\mu   \Psi^{i}_\nu - \frac{i}{2} \, \bar\Psi^i_\alpha \,  \gamma_\mu  \overset{\leftrightarrow}{D}_\nu \Psi^{i\alpha}
\nonumber\\
& + & \, \frac{i}{2}   \biggl(  \bar\Psi^i_\mu \,  \gamma_\nu  \overset{\leftrightarrow}{D}_\alpha \Psi^{i\alpha} + \bar\Psi^{i\alpha} \,  \gamma_\nu \overset{\leftrightarrow}{D}_\alpha \Psi^{i}_\mu   -  \bar\Psi^i_\mu \,  \gamma_\nu \gamma^\alpha\gamma_\beta  \overset{\leftrightarrow}{D}_\alpha \Psi^{i,\beta}  - \bar\Psi^i_\alpha \gamma^\alpha \gamma_\nu \gamma^\beta  \overset{\leftrightarrow}{D}_\mu \Psi^{i}_\beta
- \bar\Psi^i_\alpha \gamma^\alpha\gamma^\beta \gamma_\nu  \overset{\leftrightarrow}{D}_\beta \Psi^{i}_\mu  \biggl)
\nonumber\\
&+& \frac{i}{4}\,\partial^\lambda \biggl[\bar\Psi^{i, \alpha}  \biggl( \gamma_\mu \eta_{\lambda [\alpha}\eta_{\beta] \mu} +\eta_{\lambda \mu}\eta_{\nu [\alpha}\gamma_{\beta]} + \eta_{\mu\nu} \eta_{\lambda[\beta}\gamma_{\alpha]} \biggr)
\Psi^{i,\beta}   
\biggr] - \frac{m_\Delta}{2} \, \left( \bar\Psi^i_\mu \,  \gamma_\nu \gamma^\alpha \Psi^{i}_\alpha + \bar\Psi^i_\alpha \,  \gamma^\alpha \gamma_\nu \Psi^{i}_\mu \right)  
\nonumber\\
&-&  \frac{g_1}{4} \left[ 2 \bar{\Psi}^i_{\mu} u_\alpha \gamma^\alpha \gamma_5 \Psi^i_{\nu} +  \bar{\Psi}^{i,\alpha}  u_\mu  \gamma_\nu \gamma_5 \Psi^i_{\alpha}  \right]
-\frac{g_2}{4} \left[  2 \bar{\Psi}^i_\mu u_\nu  \gamma^\alpha \gamma_5 \Psi^i_{\alpha}  + 2 \bar{\Psi}^i_\alpha u_\nu  \gamma^\alpha \gamma_5 \Psi^i_{\mu}  \right.   \nonumber \\ 
    &+&  \left. \bar{\Psi}^{i,\alpha} u_\alpha  \gamma_\nu \gamma_5 \Psi^i_{\mu} + \bar{\Psi}^{i}_\mu u_\alpha  \gamma_\nu \gamma_5 \Psi^{i \alpha} \right] - \frac{g_3}{4} \left[ \bar{\Psi}^i_{\mu} u_\alpha \gamma_\nu \gamma^\alpha \gamma_5  \gamma^\beta \Psi^i_{\beta} 
      +\bar{\Psi}^i_{\beta} u_\alpha  \gamma^\beta \gamma^\alpha \gamma_5  \gamma_\nu \Psi^i_{\mu} \right.  
      \nonumber \\ &+& \left. \bar{\Psi}^i_{\alpha} u_\mu  \gamma^\alpha \gamma_\nu \gamma_5  \gamma^\beta \Psi^i_{\beta} \right]
+ \frac{\eta_{\mu\nu} }{2} \biggl[  
     \, \Bar{\Psi}^{i}_\alpha  \,  i \gamma^\beta \overset{\leftrightarrow}{D}_\beta  \Psi^{i\alpha } -  m_\Delta \, \Bar{\Psi}^{i}_\alpha   \Psi^{i\alpha}  -\Bar{\Psi}^{i}_\alpha  i \gamma^{\alpha}{\overset{\leftrightarrow}{D}_\beta}  \Psi^{i\beta} -\Bar{\Psi}^{i\alpha}i \gamma^{\beta}{\overset{\leftrightarrow}{D}_\alpha}  \Psi^{i}_\beta 
 \nonumber\\ &+&  i  \Bar{\Psi}^{i}_\rho \gamma^\rho \gamma^\alpha\gamma^\lambda \overset{\leftrightarrow}{D}_\alpha \Psi^{i}_\lambda + m_\Delta \Bar{\Psi}^{i}_\alpha \gamma^\alpha \gamma^\beta  \Psi^{i}_\beta  + 
   \frac{g_1}{2} \, \bar{\Psi}^i_{\beta} u_\alpha  \gamma^\alpha \gamma_5 \Psi^{i\beta} +  \frac{g_2}{2} \bar{\Psi}^{i\alpha}   \left( u_\alpha  \gamma_\beta +u_\beta   \gamma_\alpha  \right) \gamma_5 \Psi^{i\beta}    \nonumber \\
   &  + & \frac{g_3}{2} \bar{\Psi}^i_{\alpha}   u_\beta  \gamma^\alpha \gamma^\beta \gamma_5  \gamma^\lambda \Psi^i_{\lambda} \biggr] 
   +  \left( \mu \leftrightarrow \nu \right)  \,,
% a\frac{ i h_7}{2m} \left[ \partial^\lambda\partial_\lambda \left(\bar{\Psi}^{i\alpha} \gamma_\alpha \overset{\rightarrow}{D}_\mu \Psi^{i}_\nu - \bar{\Psi}_\nu\gamma_\alpha \overset{\leftarrow}{D}_\mu \Psi^{i\alpha}   \right) + \eta_{\mu \nu } \partial^{\kappa}\partial^{\beta} \left(\bar{\Psi}^{i\alpha} \gamma_\alpha \overset{\rightarrow}{D}_\beta\Psi^{i}_\kappa-\bar{\Psi}^{i}_\kappa \gamma_\alpha\overset{\leftarrow}{D}_\beta \Psi^{i\alpha}\right)  \right. \nonumber \\ &-& \left.  \partial^{\lambda}\partial_{\mu} \left(\bar{\Psi}^{i\alpha} \gamma_{\alpha}\overset{\rightarrow}{D}_{(\lambda} \Psi^{i}_{\nu)}-\bar{\Psi}^i_{(\nu} \gamma_{\alpha}\overset{\leftarrow}{D}_{\lambda)} \Psi^{i\alpha}\right) \right]a
\label{DeltaEMT}
\\
 T^{(1)}_{\pi N\Delta, \mu\nu}  &=& \frac{1}{2} g_{\pi N\Delta}~\eta_{\mu \nu} \left[ \bar \Psi^i_\alpha u^{\alpha}_i\Psi + \bar \Psi u^{\alpha}_i   \Psi^i_\alpha - \bar \Psi^i_\alpha \gamma^\alpha \gamma^\beta u_{\beta }^i \Psi - \bar \Psi  \gamma^\beta  \gamma^\alpha u_{\beta }^i  \Psi^i_\alpha \right] - g_{\pi N\Delta}~ \left( \bar\Psi^i_\mu u_{\nu}^i\Psi + \bar\Psi u_{\nu}^i \Psi^i_\mu  \right)  \nonumber
\\
&+&  \frac{1}{2} g_{\pi N\Delta}~ \left[ \bar \Psi^i_\mu \gamma_\nu  \gamma^\alpha u_{\alpha }^i \Psi + \bar \Psi^i_\alpha \gamma^\alpha  \gamma_\mu u_{\nu}^i \Psi + \bar \Psi  \gamma^\alpha \gamma_\nu  u_{\alpha }^i  \Psi^i_\mu + \bar \Psi  \gamma_\mu  \gamma^\alpha u_{\nu}^i  \Psi^i_\alpha \right] +   \left( \mu \leftrightarrow \nu \right),
\\
 T_{\pi \Delta,a,\mu\nu}^{(2)} & = &  a_1~\bar{\Psi}^i_{\mu} \left< \chi_+ \right> \Psi^i_{\nu} + \frac{\tilde z}{2}a_1~ \left(   \bar{\Psi}^i_{\mu} \gamma_\nu \gamma^\alpha \left< \chi_+ \right>  \Psi^i_{\alpha}   +   \bar{\Psi}^i_{\alpha} \gamma^\alpha \gamma_\mu \left< \chi_+ \right>  \Psi^i_{\nu}  \right) - \frac{a_1~}{2}\eta_{\mu \nu} \left[ \bar{\Psi}^i_{\alpha}  \left< \chi_+ \right>  \Psi^{i\alpha} \nonumber \right. \\ &+& \left. \tilde z ~ \bar{\Psi}^i_{\alpha}  \gamma^\alpha \gamma^\beta \left< \chi_+ \right>  \Psi^{i}_\beta \right] +   \left( \mu \leftrightarrow \nu \right),
\end{eqnarray}
where $A^{[\alpha} B^{\beta]} = A^{\alpha} B^{\beta} -  A^{\beta}B^{\alpha}, A^{(\alpha} B^{\beta)}
= A^{\alpha} B^{\beta} +  A^{\beta}B^{\alpha}$,  $\tilde z = z + z^\prime + n z z^\prime$ and $n$ is
spacetime dimension. 

The action of Eq.~(\ref{MActionh}) leads to the following expression for the EMT  in flat
spacetime (we  have dropped terms involving $h_2, h_3, h_7, h_8, h_9, h_{14}$ and $h_{15}$,
because they do not give any contributions in our analysis of the delta GFFs):

\begin{eqnarray}
 T^{(2)}_{\pi\Delta,b,\mu\nu}  & = &   h_1  \left(\eta_{\mu \nu} \partial_\lambda\partial^\lambda - \partial_{\mu}\partial_{\nu}\right)\bar{ \Psi}^i_\alpha \Psi^{i\alpha}   +   \frac{h_4}{2} \left[ \partial^\lambda\partial_\lambda  \left(\bar{ \Psi}^i_\nu  \Psi^{i}_\mu \right) +  \eta_{\mu\nu} \partial^\alpha \partial^\beta \left( \bar{ \Psi}^i_\beta \Psi^{i}_\alpha\right)  \right. \nonumber \\  &-&  \left.  \partial^\lambda \partial_\mu \left(   \bar{ \Psi}^i_{(\lambda} \Psi^{i}_{\nu )} \right)\ \right] +  i h_5  \left[ \partial^\lambda\partial_\lambda \left(\bar{ \Psi}^i_\alpha \gamma_\mu \overset{\leftrightarrow}{D}_\nu \Psi^{i\alpha}\right)  + \eta_{\mu \nu } \partial^{\kappa}\partial^{\beta} \left(\bar{ \Psi}^i_\alpha \gamma_\beta \overset{\leftrightarrow}{D}_\kappa \Psi^{i\alpha}\right)- \partial^{\lambda}\partial_{\mu} \left(\bar{ \Psi}^i_\alpha \gamma_{(\lambda} \overset{\leftrightarrow}{D}_{\nu)} \Psi^{i\alpha}\right) \right]  \nonumber \\  &+&   \frac{ i h_6}{2} \left[ \partial^\lambda\partial_\lambda \left(\bar{ \Psi}^{i\alpha} \gamma_\mu \overset{\rightarrow}{D}_\alpha  \Psi^{i}_\nu - \bar{ \Psi}_\nu\gamma_\mu \overset{\leftarrow}{D}_\alpha  \Psi^{i\alpha}   \right) + \eta_{\mu \nu } \partial^{\kappa}\partial^{\beta} \left(\bar{ \Psi}^{i\alpha} \gamma_\beta \overset{\rightarrow}{D}_\alpha \Psi^{i}_\kappa-\bar{ \Psi}^{i}_\kappa \gamma_\beta \overset{\leftarrow}{D}_\alpha \Psi^{i\alpha}\right) \right. \nonumber \\ &-& \left.  \partial^{\lambda}\partial_{\mu} \left(\bar{ \Psi}^{i\alpha} \gamma_{(\lambda} \overset{\rightarrow}{D}_{\alpha} \Psi^{i}_{\nu)}-\bar{ \Psi}^i_{(\nu} \gamma_{\lambda)} \overset{\leftarrow}{D}_{\alpha} \Psi^{i\alpha}\right) \right] + i h_{10} ~ \partial^{\kappa}\partial^{\beta} \left(  \bar{ \Psi}^{i}_\kappa \sigma_{\beta \nu}  \Psi^{i}_{\mu}  -   \bar{ \Psi}^{i}_{\mu} \sigma_{\beta \nu}  \Psi^{i}_\kappa  \right)  \nonumber \\ &+& \frac{ i h_{11}}{2} \partial^{\kappa} \partial^{\beta} \left[\bar{ \Psi}^i_\kappa \gamma_\beta  \overset{\rightarrow}{D}_\mu  \Psi^i_\nu-\bar{ \Psi}^i_\kappa \gamma_\nu  \overset{\rightarrow}{D}_\beta  \Psi^i_\mu +  \bar{ \Psi}^i_\mu \gamma_\nu  \overset{\rightarrow}{D}_\beta  \Psi^i_\kappa -   \bar{ \Psi}^i_\nu \gamma_\beta  \overset{\rightarrow}{D}_\mu  \Psi^i_\kappa -\bar{ \Psi}^i_\nu \gamma_\beta  \overset{\leftarrow}{D}_\mu  \Psi^i_\kappa    + \bar{ \Psi}^i_\mu \gamma_\nu  \overset{\leftarrow}{D}_\beta  \Psi^i_\kappa \right. \nonumber \\ &-& \left. \bar{ \Psi}^i_\kappa \gamma_\nu  \overset{\leftarrow}{D}_\beta  \Psi^i_\mu+\bar{ \Psi}^i_\kappa \gamma_\beta  \overset{\leftarrow}{D}_\mu  \Psi^i_\nu \right] + \frac{ i h_{12}}{2} \partial^{\kappa} \partial^{\beta} \left[\bar{ \Psi}^i_\mu \gamma_\beta  \overset{\rightarrow}{D}_\kappa  \Psi^i_\nu - \bar{ \Psi}^i_\beta \gamma_\nu  \overset{\rightarrow}{D}_\kappa  \Psi^i_\mu +\bar{ \Psi}^i_\beta \gamma_\nu  \overset{\rightarrow}{D}_\mu  \Psi^i_\kappa  -  \bar{ \Psi}^i_\mu \gamma_\beta  \overset{\rightarrow}{D}_\nu  \Psi^i_\kappa  \right. \nonumber \\ &-& \left.    \bar{ \Psi}^i_\nu \gamma_\beta  \overset{\leftarrow}{D}_\kappa   \Psi^i_\mu + \bar{ \Psi}^i_\mu \gamma_\nu  \overset{\leftarrow}{D}_\kappa  \Psi^i_\beta-  \bar{ \Psi}^i_\kappa \gamma_\nu  \overset{\leftarrow}{D}_\mu  \Psi^i_\beta + \bar{ \Psi}^i_\kappa \gamma_\beta  \overset{\leftarrow}{D}_\nu  \Psi^i_\mu \right] + \frac{  h_{13}}{2} \partial^{\kappa} \partial^{\beta} \left[ \eta_{\mu \nu } \bar{ \Psi}^i_\beta  \Psi^i_\kappa - \bar{ \Psi}^i_\beta \gamma_\nu \gamma_\kappa  \Psi^i_\mu \right. \nonumber \\ &-& \left.   \bar{ \Psi}^i_\mu \gamma_\beta\gamma_{\nu}   \Psi^i_\kappa  + \bar{ \Psi}^i_\mu \gamma_\beta \gamma_\kappa  \Psi^i_\nu  \right] + \left( \mu \leftrightarrow \nu \right)  \,.
% a\frac{ i h_7}{2m} \left[ \partial^\lambda\partial_\lambda \left(\bar{\Psi}^{i\alpha} \gamma_\alpha \overset{\rightarrow}{D}_\mu \Psi^{i}_\nu - \bar{\Psi}_\nu\gamma_\alpha \overset{\leftarrow}{D}_\mu \Psi^{i\alpha}   \right) + \eta_{\mu \nu } \partial^{\kappa}\partial^{\beta} \left(\bar{\Psi}^{i\alpha} \gamma_\alpha \overset{\rightarrow}{D}_\beta\Psi^{i}_\kappa-\bar{\Psi}^{i}_\kappa \gamma_\alpha\overset{\leftarrow}{D}_\beta \Psi^{i\alpha}\right)  \right. \nonumber \\ &-& \left.  \partial^{\lambda}\partial_{\mu} \left(\bar{\Psi}^{i\alpha} \gamma_{\alpha}\overset{\rightarrow}{D}_{(\lambda} \Psi^{i}_{\nu)}-\bar{\Psi}^i_{(\nu} \gamma_{\alpha}\overset{\leftarrow}{D}_{\lambda)} \Psi^{i\alpha}\right) \right]a
\label{DeltaEMTH}
\end{eqnarray}
The covariant derivatives ${D}$ acting on spin-1/2 and spin-3/2 fields in $T^{{  (\pi N)}}_{\mu\nu} $,
$T^{{  (\pi \Delta)}}_{\mu\nu} $ and $T^{{  (\Delta)}}_{\mu\nu} $ coincide with $\nabla$ in Eq.~(\ref{CovD})
with $g_{\mu \nu} = \eta_{\mu\nu}$, i.e. $ \Gamma_{\mu\nu}^\beta = \omega_{\mu}^{ab} = 0$.

 The above expressions of the EMT can be used for the calculations of various matrix elements
between states containing one nucleon and/or delta resonance
and an arbitrary number of pions at low energies. Below we consider the corrections to the
GFFs of the delta resonance at leading one-loop order.

\begin{figure}[t]
\begin{center}
\epsfig{file=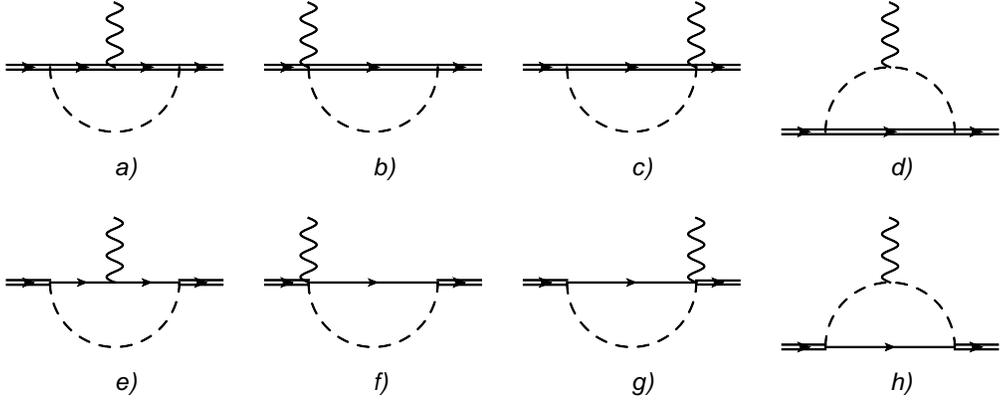,scale=0.8}
\caption{One-loop  diagrams contributing to the one-particle matrix elements of the EMT for the
delta resonances. Dashed, solid and double lines correspond to pions, nucleons and delta
resonances, respectively. The wiggly line indicates the EMT insertion. 
%\tbd{I think, using a more standard notation for the delta field with double lines would make the figure better readable. Otherwise, the lines can easily be confused.}
} 
%{\color{magenta} J.G.: These figures are drawn using Jaxodraw. As far as I can see this program does not offer an option of double lines, otherwise I would be happy to redraw.  }
%The circles with crosses represent the EMT vertices. 
\label{EMT_delta}
\end{center}
\vspace{-5mm}
\end{figure}

\section{{One-loop corrections to the gravitational form factors of the delta resonances}}
\label{NFFs}

In this section we calculate leading one-loop contributions to the matrix elements of the
EMT for the one-particle states of the delta resonance 
We extract these matrix elements from the residues of Green's functions at complex
poles of the initial and final four-momenta squared,
corresponding to the unstable delta-states \cite{Gegelia:2010nmt}.

The one-loop diagrams contributing to our calculations are shown in Fig.~\ref{EMT_delta}. 
We apply the so called $\epsilon$-counting scheme (also called the small scale
expansion)\footnote{For an alternative power counting in an EFT with delta resonances, see Ref.~\cite{Pascalutsa:2002pi}.}, 
i.e.~the pion lines count as of chiral order minus two, the nucleon and delta lines
have order minus one,   interaction vertices originating from the effective Lagrangian 
of order $N$ count also as of chiral order $N$ and the vertices generated by the EMT
have the orders corresponding 
to the number of quark mass factors and derivatives acting on the pion fields. 
Derivatives acting on the nucleon and delta fields count as of chiral order 
zero. The momentum transfer between the initial and final states of the delta resonance
also counts as of chiral order one, 
therefore in those terms of EMT which contain full derivatives, these derivatives
count as of chiral order one.  
Integration over loop momenta is counted as of chiral order four. The delta-nucleon
mass difference also counts as of order one within the $\epsilon$-counting scheme.

Since we are interested in the delta matrix elements of order three in the chiral expansion,
we need vertices with two delta lines, generated by  the EMT, up to third order. 
From the effective Lagrangian specified above, we have obtained these vertices from the
expressions of the EMT for the zeroth, first and second chiral orders, while there are no
tree-order contributions at third order.  
Simple power counting arguments show that for all one-loop diagrams except f)  
we only need vertices up to order one. Naively it seems that for diagrams of
topology a) and f) we need also pion-baryon-baryon vertices of chiral order two,
because the gravitational-souce-baryon-baryon vertices originating from the  EMT starts
with chiral order zero. However, the leading contribution of the diagrams with the
mentioned zeroth order  vertices is exactly canceled by the wave function renormalization
constant of the delta resonance multiplying the tree-order diagrams. Therefore, the formally
zeroth order vertices in effect  start contributing as vertices of order one.  As a result, the
diagrams with the pion-baryon-baryon vertex of order two only start  contributing at chiral
order four. For this reason we do not consider such diagrams in this work. It is understood
that the above described power counting for loop diagrams is realized as the result 
of our manifestly Lorentz-invariant calculations only after performing an appropriate
renormalization.  To get rid off the divergent parts and power counting violating pieces
from the expressions of one-loop diagrams we apply the  EOMS scheme of
Refs.~\cite{Gegelia:1999gf,Fuchs:2003qc}.

\subsection{Gravitational form factors of the delta resonance}

The matrix element of the total EMT for the delta resonances can be parameterized in
terms of seven form factors \cite{Cotogno:2019vjb,Kim:2020lrs}\footnote{Straightforward calculation
of the matrix elements in Eq.~(\ref{eqME}) within chiral EFT leads to nine invariant structures.
It can, however, be shown that only seven structures are independent \cite{Cotogno:2019vjb}.
The two redundant structures can be eliminated using on-shell identities as given in
Refs.~\cite{Cotogno:2019vjb} and \cite{Fu:2022rkn}.}:
\begin{eqnarray}
\langle p_f, s_f| T_{\mu\nu}  | p_i,s_i \rangle & = & - \bar u_{\alpha'}(p_f, s_f) \left[ \frac{ P^{\mu}P^{\nu}}{m} \left(\eta^{{\alpha^\prime}\alpha}F_{1,0}(t) -\frac{\Delta^{{\alpha^\prime}}\Delta^\alpha}{2 m_\Delta^2} F_{1,1}(t)\right)\right. \nonumber \\ &+&  \frac{\Delta^{\mu}\Delta^{\nu} - \eta^{\mu\nu} \Delta^2}{4m} \left(\eta^{{\alpha^\prime}\alpha}F_{2,0}(t) -\frac{\Delta^{{\alpha^\prime}}\Delta^\alpha}{2 m_\Delta^2} F_{2,1}(t)\right) \nonumber \\ \ &+&  \frac{i}{2 m_\Delta} P^{ \{ \mu} \sigma^{ \nu \} \rho}  \Delta_{\rho} \left(\eta^{{\alpha^\prime}\alpha}F_{4,0}(t) -\frac{\Delta^{{\alpha^\prime}}\Delta^\alpha}{2 m_\Delta^2} F_{4,1}(t)\right) \nonumber \\ &-& \left.\frac{1}{m_\Delta} \left( \eta^{\alpha\{\mu} \Delta^{\nu\}} \Delta^{\alpha^\prime}  + \eta^{\alpha^\prime \{\mu} \Delta^{\nu\}} \Delta^{\alpha} - 2 \eta^{\mu\nu} \Delta^{\alpha}\Delta^{\alpha^\prime} - \Delta^2 \eta^{\alpha \{ \mu} \eta^{\nu\} \alpha^\prime} \right)F_{5,0}(t)\right] u_\alpha(p_i, s_i) ,
\label{eqME}
\end{eqnarray} 
where $m_\Delta$ is the physical mass of the delta resonances (we work in isospin symmetric limit),
$(p_i,s_i)$ and $(p_f,s_f)$ are the momenta and polarizations of the incoming and outgoing particles,
respectively, and $P=(p_i+p_f)/2$, $\Delta=p_f-p_i$, { $t=\Delta^2$}.  

The tree-order diagrams contributing to the matrix element of the EMT up to third chiral
order yield the following contributions to the form factors:
\begin{eqnarray}
F_{\rm 1,0, tree}(t) &=&  1 -\frac{t}{m_\Delta^2} + \frac{t \left(2 h_5 m_{\Delta }+2 h_{10}-h_{13}\right)}{m_{\Delta
   }}-\frac{\left(-2 h_6+2 h_{11}+h_{12}\right) t^2}{2 m_{\Delta }^2} \,,\nonumber\\
F_{\rm 1,1, tree}(t) &=&   -4 - 4 m_{\Delta } \left(h_{12}
m_{\Delta }-2 h_{10}+h_{13}\right) + \left(4 h_6-2 \left(2 h_{11}+h_{12}\right)\right) t  \,,\nonumber
\end{eqnarray}
\begin{eqnarray}
F_{\rm 2,0, tree}(t) &=&  -2 -4 \left(2 h_1-2 h_{10}+h_{13}\right)
   m_{\Delta } + \left(2 h_6-2 h_{11}-h_{12}\right) t  \,,\nonumber\\
F_{\rm 2,1, tree}(t) &=&  0 \,,\nonumber\\
F_{\rm 4,0, tree}(t) &=&  \frac{3}{2}  -\frac{t}{2 m_\Delta^2} +t \left(\frac{h_{10}}{m_{\Delta }}-\frac{h_{13}}{2 m_{\Delta
   }}+h_5-h_6+h_{11}+\frac{h_{12}}{2}\right)-\frac{\left(-2 h_6+2
   h_{11}+h_{12}\right) t^2}{4 m_{\Delta }^2} \,,\nonumber\\
F_{\rm 4,1, tree}(t) &=&  -2 - 2 m_{\Delta } \left(h_{12} m_{\Delta }-2
   h_{10}+h_{13}\right) +\left(2 h_6-2 h_{11}-h_{12}\right) t  \,,\nonumber\\
F_{\rm 5,0, tree}(t) &=&  -\frac{1}{2} +\frac{1}{2} \left(h_4+4 h_{10}-h_{13}\right) m_{\Delta }+\frac{1}{4} \left(2
   h_6-2 h_{11}-h_{12}\right) t \,,
\label{temt}
\end{eqnarray}
where the $h_i$-terms are generated by the EMT of Eq.~(\ref{DeltaEMTH}).

In the calculations of loop diagrams shown in Fig.~\ref{EMT_delta},  we apply dimensional
regularization (see, e.g., Ref.~\cite{Collins:1984xc}) and use the program
FeynCalc \cite{Mertig:1990an,Shtabovenko:2016sxi}. The one-loop expressions of the form
factors are too lengthy to be shown explicitly. They are available from 
the authors upon request.

To get rid of the power-counting violating contributions we split the bare low-energy
parameters as the renormalized ones and counterterms. 
We specify the finite parts of counterterms by applying the EOMS scheme with the remaining
renormalization scale chosen as $\mu=m_N$, where  $m_N$ is the mass of the nucleon. 
The one-loop finite parts of counter terms $\delta h_i$  are given by:
\begin{eqnarray}
\delta h_1 & = & \frac{\delta h_{12} m_N}{2}-\frac{\left(1575 \,g_{\pi N\Delta}^2+172\,
g_1^2\right) m_N}{207360 \pi ^2 F^2} \,,\nonumber\\
\delta h_4 & = & -2 \,\delta h_{10} - \delta h_{12} m_N - \frac{m_N(45 \,g_{\pi N\Delta}^2+2336 \,
g_1^2)}{51840 \pi ^2 F^2} \,,\nonumber\\
\delta h_5 & = & -\frac{\delta h_{12}}{2}  - \frac{11( 135 \,g_{\pi N\Delta}^2 +124\, g_1^2)}
{207360 \pi ^2 F^2}  \,,\nonumber\\
\delta h_{13} & = & 2 \, \delta h_{10} - \delta h_{12}
m_N +\frac{\left(9 \, g_{\pi N\Delta}^2+490 \, g_1^2\right) m_N}{10368 \pi ^2 F^2} \,.
\label{counterterms}
\end{eqnarray}

After renormalization, we obtain the following expressions for the GFFs at $t=0$, expanded in powers of the pion mass 
and the delta-nucleon mass difference in the chiral limit $\delta$:
\begin{eqnarray}
F_{\rm 1,0, loop}(0) &=&    0  \,,\nonumber\\
F_{\rm 1,1, loop}(0) &=&   % \frac{g^2 m_N^2}{288 \pi ^2 F^2}+\frac{245 g_1^2 m_N^2}{1296 \pi ^2 F^2} 
   -\frac{5 g_1^2 m_N (3 \pi  M-49 \delta )}{648 \pi ^2 F^2} \nonumber\\
   &+& \frac{g_{\pi N\Delta}^2  m_N }{144 \pi
   ^2 F^2 \left(M^2-\delta ^2\right)}
   \Biggl( -53 \delta ^3+24 \delta  \left(M^2-\delta ^2\right)
   \ln  \frac{M}{m_N} +24 i \pi  \delta ^2 \sqrt{\delta
   ^2-M^2}-12 i \pi  M^2 \sqrt{\delta ^2-M^2} \nonumber \\
   &+& 12 \left(M^2-2 \delta
   ^2\right) \sqrt{\delta ^2-M^2} \ln \frac{\delta
   +\sqrt{\delta ^2-M^2}}{M} +53 \delta  M^2 \Biggr) 
    +{\cal O} (\epsilon^2)   \,,\nonumber\\
F_{\rm 2,0, loop}(0) &=&  % \frac{197 g_1^2 m_N^2}{1080 \pi ^2 F^2}-\frac{11 g^2 m_N^2}{192 \pi ^2 F^2}  
   -\frac{g_1^2 m_N (25 \pi  M-1068 \delta )}{2160 \pi ^2
   F^2} \nonumber\\
   &+& \frac{g_{\pi N\Delta}^2 m_N \left(29 \delta +48 \delta  \ln
    \frac{M}{m_N} -48 i \pi  \sqrt{\delta ^2-M^2}+48
   \sqrt{\delta ^2-M^2} \ln \frac{\delta +\sqrt{\delta
   ^2-M^2}}{M}\right)}{288 \pi ^2 F^2} +{\cal O} (\epsilon^2)   \,,\nonumber\\
F_{\rm 2,1, loop}(0) &=&  -\frac{g_1^2 m_N^3}{54 \pi  F^2 M}+\frac{g_{\pi N\Delta}^2 M m_N^3
   \sqrt{\frac{\delta ^2}{M^2}-1} \left(\ln
   \left(\sqrt{\frac{\delta ^2}{M^2}-1}+\frac{\delta }{M}\right)-i
   \pi \right)}{15 \pi ^2 F^2 \left(M^2-\delta ^2\right)}  +{\cal O} (\epsilon^0)    \,,\nonumber\\
F_{\rm 4,0, loop}(0) &=&  0    \,,\nonumber\\
%\end{eqnarray}
%\begin{eqnarray}
F_{\rm 4,1, loop}(0) &=&   \frac{5 g_{\pi N\Delta}^2 m_N^2}{576 \pi ^2 F^2}+\frac{235 g_1^2 m_N^2}{2592 \pi
   ^2 F^2}   +{\cal O} (\epsilon)  \,,\nonumber\\
F_{\rm 5,0, loop}(0) &=& % \frac{g^2 m_N^2}{1152 \pi ^2 F^2}+\frac{2393 g_1^2 m_N^2}{51840 \pi ^2 F^2}  
   -\frac{g_1^2 m_N (150 \pi  M-3323 \delta )}{25920 \pi ^2
   F^2} \nonumber\\
   &+& \frac{g_{\pi N\Delta}^2 m_N \left(5 \delta +2 \delta  \ln
   \frac{M}{m_N}  -2 i \pi  \sqrt{\delta ^2-M^2}+2
   \sqrt{\delta ^2-M^2} \ln \frac{\delta +\sqrt{\delta
   ^2-M^2}}{M}\right)}{96 \pi ^2 F^2}   +{\cal O} (\epsilon^2)    \,.
\label{FFs0}
\end{eqnarray}

Next, we define the slopes of the GFFs by writing the form factors as:
\begin{eqnarray}
F_{i,j}(t) & = & F_{i,j}(0) + s_{F_{i,j}} t +{\cal O}(t^2) \,.
\label{defradii}
\end{eqnarray}
Calculating loop contributions to these quantities and expanding in powers of the pion mass and $\delta$ we obtain
\begin{eqnarray}
s_{F_{1,0}} & = &   
%  \frac{35 g^2}{2304 \pi ^2 F^2}+\frac{29 g_1^2}{480 \pi ^2 F^2} + 
 \frac{g_1^2 (8 \delta -255 \pi  M)}{10368 \pi ^2 F^2 m_N}
 \nonumber\\
 &+& \frac{g_{\pi N\Delta}^2 }{576 \pi
   ^2 F^2 m_N \left(M^2-\delta ^2\right)}  
   \Biggl( 25 \delta( \delta ^2 - M^2)  +24 \delta  \left(\delta ^2-M^2\right) \ln
    \frac{M}{m_N} - 12 i \pi  ( 2 \delta ^2 -M^2) \sqrt{\delta
   ^2-M^2}
   \nonumber\\
   & - & 12 \left(M^2-2 \delta
   ^2\right) \sqrt{\delta ^2-M^2} \ln \frac{\delta
   +\sqrt{\delta ^2-M^2}}{M}\Biggr)  + {\cal O} (\epsilon^2) \,,\nonumber\\
 s_{F_{1,1}} & = &  \frac{g_1^2 m_N}{432 \pi  F^2 M}+\frac{g_{\pi N\Delta}^2 m_N \left(\delta ^3+M^2
   \left(-\delta +i \pi  \sqrt{\delta ^2-M^2}\right)-M^2
   \sqrt{\delta ^2-M^2} \ln \frac{\delta +\sqrt{\delta
   ^2-M^2}}{M} \right)}{120 \pi ^2 F^2 \left(M^2-\delta
   ^2\right)^2} 
   +{\cal O} (\epsilon^0) \,,\nonumber\\
 s_{F_{2,0}} & = &  -\frac{g_1^2 m_N}{108 \pi  F^2 M}+\frac{g_{\pi N\Delta}^2 m_N \left(\ln
    \frac{\delta +\sqrt{\delta ^2-M^2}}{M} -i \pi
   \right)}{60 \pi ^2 F^2 \sqrt{\delta ^2-M^2}} 
   +{\cal O} (\epsilon^0) \,,\nonumber\\
 s_{F_{2,1}} & = &\frac{g_{\pi N\Delta}^2 m_N^3 \left(-\delta
   ^3+M^2 \left(\delta -i \pi  \sqrt{\delta ^2-M^2}\right)+M^2
   \sqrt{\delta ^2-M^2} \ln  \frac{\delta +\sqrt{\delta
   ^2-M^2}}{M} \right)}{140 \pi ^2 F^2 M^2 \left(M^2-\delta
   ^2\right)^2} -\frac{g_1^2 m_N^3}{504 \pi  F^2 M^3} +{\cal O} (\epsilon^{-2}) \,,\nonumber\\
 s_{F_{4,0}} & = &  \frac{g_{\pi N\Delta}^2
   \left(163 \delta ^2-96 \left(M^2-\delta ^2\right) \ln
   \frac{M}{m_N}-96 i \pi  \delta  \sqrt{\delta
   ^2-M^2}+96 \delta  \sqrt{\delta ^2-M^2} \ln \frac{\delta
   +\sqrt{\delta ^2-M^2}}{M} -163 M^2\right)}{4608 \pi ^2 F^2
   \left(M^2-\delta ^2\right)} \nonumber\\
   & + & \frac{g_1^2 \left(877-150 \ln
    \frac{M}{m_N} \right)}{25920 \pi ^2 F^2}
   +{\cal O} (\epsilon) \,,\nonumber\\
 s_{F_{4,1}} & = &   
 0 +{\cal O} (\epsilon^{-1}) \,,\nonumber\\
 s_{F_{5,0}} & = &\frac{g_1^2 m_N}{3456 \pi  F^2 M}+\frac{g_{\pi N\Delta}^2 m_N \left(\ln
   \frac{\delta +\sqrt{\delta ^2-M^2}}{M}  -i \pi
   \right)}{960 \pi ^2 F^2 \sqrt{\delta ^2-M^2}}  +{\cal O} (\epsilon^0) \,.
\label{radii}
\end{eqnarray}
%\tbd{Why is only the leading term ($0$) and not the subleading one shown for $F_{\rm 4,1, loop}(0)$ as for the other slopes?} {\color{magenta} 
%All slopes are shown only up to the order to which the results of the matrix element up to third order give contributions. Subleading order for the slope of $F_{\rm 4,1, loop}(0)$ gets contributions from fourth order amplitude. That is, the subleading result generated from the third order amplitude is incomplete.}
%are
Note that the tree-order contributions to slopes are included in Eq.~(\ref{temt}).
Notice further that these expressions as well as the expressions in Eq.~(\ref{FFs0}) contain
(unphysical) singularities in the $M\to\delta$ limit, which is due to effect first
uncovered, to the best of our knowledge, in
Ref.~\cite{Ledwig:2010ya} for the electromagnetic interaction. In particular, in this limit,
the one-photon exchange approximation fails completely and one needs to resum the series 
of multiphoton exchange diagrams. The same is also true for the gravitational interaction
manifested in the above mentioned singularities.

\begin{figure}[t]
\begin{center}
\epsfig{file=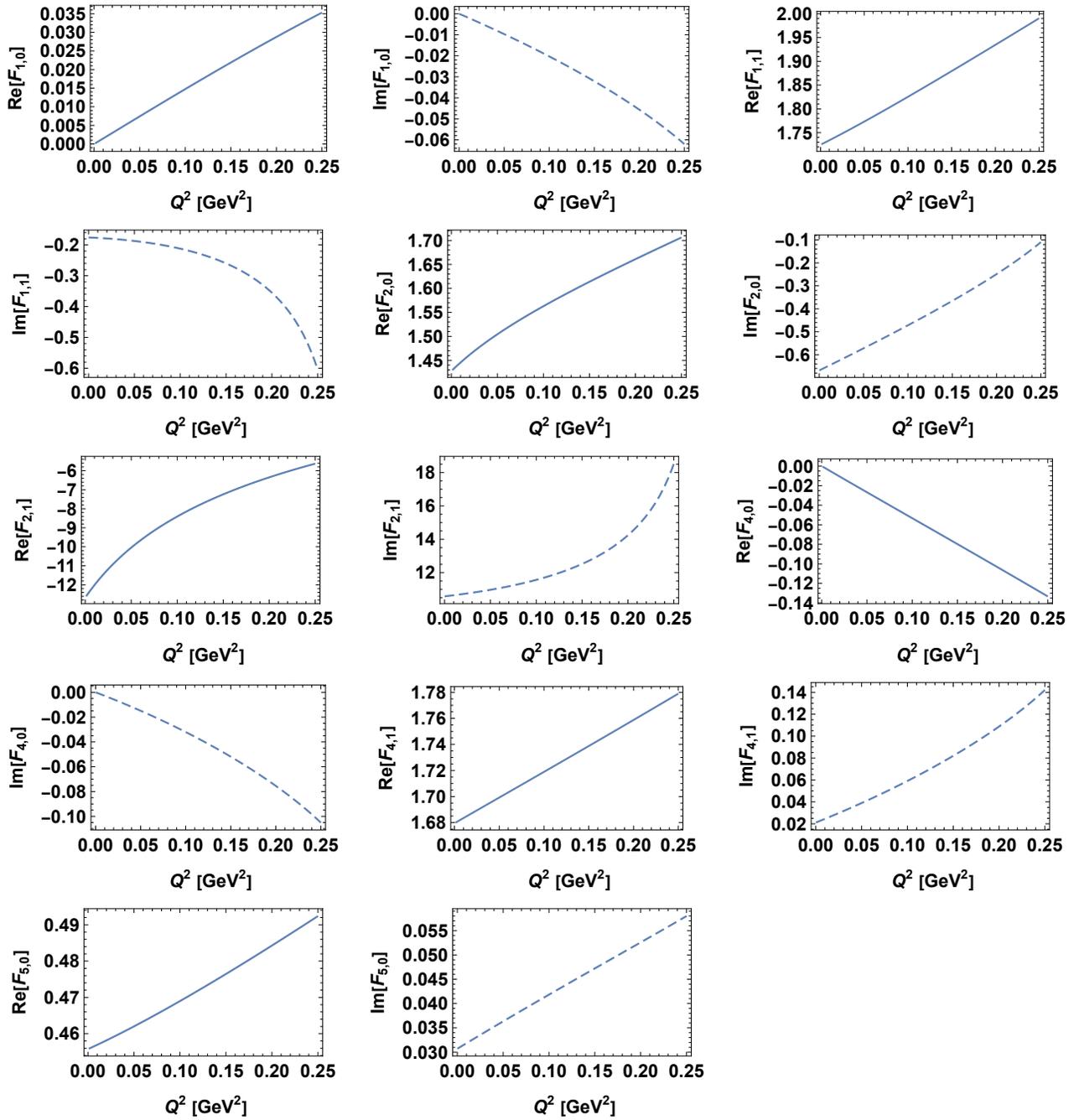,scale=0.95}
\caption{Contributions of renormalized (subtracted) one-loop  diagrams to the GFFs of the
delta resonances. Solid and dashed lines correspond to real and imaginary parts, respectively.}
\label{deltaFFs}
\end{center}
\vspace{-5mm}
\end{figure}

The full one-loop contributions to GFFs consist of real and imaginary parts also for
spacelike transfer momenta. This is because of the deltas being unstable particles. Due to
the lack of empirical data, from which we could fix the free parameters contributing at
tree order  in Fig.~\ref{deltaFFs}, we plot 
the contributions of renormalized (subtracted) one-loop  diagrams as functions of $Q^2=-t$.

\section{Summary}
\label{conclusions}

In the framework of chiral EFT for pions, nucleons and delta resonances interacting with an
external gravitational field, we calculated the leading one-loop contributions to 
the one-particle  matrix elements of the EMT for delta resonances and extracted
the corresponding  contributions to the gravitational form factors. 
To get rid of the UV divergences and power counting violating pieces from the loop diagrams
we applied the EOMS renormalization scheme of Refs.~\cite{Gegelia:1999gf,Fuchs:2003qc}.
Since the delta resonances are unstable particles, the loop contributions to the gravitational
form factors are complex valued quantities also for space-like momentum transfers.   
This is manifested in contributions of one-loop diagrams to the real and imaginary parts
of the GFFs, see Fig.~\ref{deltaFFs}. Unfortunately, no empirical data are available, 
from which we could deduce the low-energy constants contributing at tree order. 
We also give analytic expressions of GFFs at zero transfers and slope parameters in
the form of expansions in small parameters. 
Notice that the value as well as the sign of the $D$-term of delta resonances cannot be
predicted/calculated  within chiral EFT.

\acknowledgements 
We are grateful to Dongyan Fu for the discussion on reducing the redundant terms in the
EMT parametrization. This work was supported in part by 
DFG and NSFC through funds provided to the Sino-German CRC 110
“Symmetries and the Emergence of Structure in QCD” (NSFC Grant
No. 11621131001, DFG Project-ID 196253076 - TRR 110),
%by ERC  NuclearTheory (grant No. 885150) and ERC EXOTIC (grant No. 101018170),
by CAS through a President’s International Fellowship Initiative (PIFI)
(Grant No. 2018DM0034), by the VolkswagenStiftung
(Grant No. 93562), and by the EU Horizon 2020 research and
innovation programme (STRONG-2020, grant agreement No. 824093), by Guangdong Provincial
funding with Grant
No. 2019QN01X172, the National Natural Science Foundation of China
with Grant No. 12035007 and No. 11947228, Guangdong Major Project of
Basic and Applied Basic Research No. 2020B0301030008, and the Department of Science and
Technology of Guangdong Province with Grant No. 2022A0505030010.

\end{document}